\documentclass[12pt, a4paper]{article}

\newcommand{\sect}[1]{ \section{#1} }

\newcommand{\ve}{\left( \begin{array}{r}}
\newcommand{\ev}{\end{array} \right)}
\newcommand{\ar}{\left( \begin{array}{rr}}
\newcommand{\ra}{\end{array} \right)}
\newcommand{\arr}{\left( \begin{array}{rrrr}}
\newcommand{\arrr}{\left( \begin{array}{rrrrrr}}

\newcommand{\eqr}{\begin{eqnarray}}
\newcommand{\rqe}{\end{eqnarray}}
\newcommand{\eq}{\begin{equation}}
\newcommand{\qe}{\end{equation}}

\def\KK{{\rm I\kern -.2em  K}}
\def\NN{{\rm I\kern -.16em N}}
\def\RR{{\rm I\kern -.2em  R}}

\def\ZZZ{{\small{\rm Z}\kern -.5em Z}}
\def\QQ{{\rm \kern .25em
             \vrule height1.4ex depth-.12ex width.06em\kern-.31em Q}}
\def\CC{{\rm \kern .25em
             \vrule height1.4ex depth-.12ex width.06em\kern-.31em C}}

\title{Supergravity M5-branes wrapped on Riemann surfaces
and their QFT duals}
\author{Bj\"orn Brinne$^1$\footnote{email: brinne@physto.se} \and
Ansar Fayyazuddin$^1$\footnote{email: ansar@physto.se} \and
Subir Mukhopadhyay$^1$\footnote{email: subir@physto.se} \and
Douglas J. Smith$^2$\footnote{email: Douglas.Smith@durham.ac.uk} }

\begin{document}

\maketitle

\begin{center}

{\em
\vspace{-0.5cm}
$^1$Department of Physics\\
Stockholm University\\
Box 6730\\
S-113 85 Stockholm\\
Sweden \\
\vspace{0.5cm}
$^2$Department of Mathematical Sciences \\
University of Durham \\
Durham \\
DH1 3LE \\
UK
}

\end{center}

\vspace{0.5cm}

\begin{abstract}
We find solutions of 11-dimensional supergravity for M5-branes
wrapped on Riemann surfaces.  These solutions preserve
${\cal N} = 2$ four-dimensional supersymmetry.  They are
dual to ${\cal N} = 2$ gauge theories, including non-conformal field
theories.  We work out the case of ${\cal N} = 2$ Yang-Mills in detail.

\end{abstract}

\vspace{-18.5cm}
\begin{flushright}
USITP-00-11 \\
DTP/00/55 \\
hep-th/0009047
\end{flushright}

\thispagestyle{empty}

\newpage

\setcounter{page}{1}

\sect{Introduction}

The AdS/CFT duality \cite{Maldacena, AdS_refs} and its generalizations relate 
supergravity in certain backgrounds to quantum field theories.
Most notably ${\cal N} = 4$ Yang-Mills has been studied
extensively through its dual AdS$_5\times S^5$ background in type IIB
supergravity.  These backgrounds arise as near-horizon geometries
of branes on which the quantum field theories live as world-volume
theories.  
Some quantum field theories with lower supersymmetry have
been studied by using a number of techniques including orbifolds
and orientifolds probed by D3-branes. Others have been constructed
through flows of the ${\cal N} = 4$ theory by supersymmetry preserving
perturbations \footnote{See \cite{bigrev} for an introduction to these 
techniques and references.}. In this paper we study duals of ${\cal N} = 2$ gauge
theories realized as M5-branes wrapped on Riemann surfaces.  

In a previous paper, two of us found a solution for a particular
Riemann surface which described a set of intersecting M5-branes.
In the present work we generalize the construction to all wrapped
M5-brane configurations which preserve 8 real supersymmetries
(corresponding to ${\cal N} = 2$ in four dimensions).  One of the
advantages of this construction is that given an ${\cal N} = 2$ gauge
theory described in terms of a Seiberg-Witten Riemann surface
one can construct a dual supergravity solution.  Thus this is a step
towards a systematic classification of geometries dual to 
${\cal N} = 2$ gauge theories.  Another approach to finding solutions
for wrapped M5-branes which can be generalized to 
arbitrary Riemann surfaces was presented in \cite{maldanunez}.  

\section{M5-brane setup}
 
We will first describe the M5-brane configurations we want to consider. Much
of this is discussed in more detail in \cite{danda} and the special case of
orthogonally intersecting branes is solved in \cite{LastPaper}.

We are interested in Hanany-Witten configurations \cite{HW} describing
${\cal N} = 2$ gauge theories in four dimensions. As shown in
\cite{vafa, witn2}, the relevant M-theory description is in terms of an M5-brane
wrapped on a (non-compact) Riemann surface $\Sigma$ which is identified with
the Seiberg-Witten curve. We want to consider the supergravity description of
this system. In particular, the near-horizon limit in supergravity is
expected to provide a supergravity dual description of the field theory. This
would give a large class of conformal and non-conformal examples of
Maldacena's AdS/CFT conjecture \cite{Maldacena} and its generalizations.

In our notation, the M5-brane worldvolume is $\RR^{3,1} \times \Sigma$ where
$\RR^{3,1}$ has coordinates $x^{\mu}$ (where $\mu$ runs from 0 to 3) and
$\Sigma$ is a holomorphic curve in $\CC^2$ with complex coordinates $v$ and
$s$. In terms of real coordinates:
\begin{eqnarray}
v \equiv z^1 & = &x^4 + ix^5\\
s \equiv z^2 & = &x^6 + ix^7. \label{coord}
\end{eqnarray}
Ten-dimensional type IIA theory is reached by compactifying $x^7$ on a circle of
radius $R$. After compactifying, the gauge theory is seen to arise as the
worldvolume theory on D4-branes suspended between NS5-branes \cite{witn2}.
The remaining coordinates $x^{\alpha}$ (where
$\alpha$ runs from 8 to 10) are transverse to the M5-brane. The supergravity
solution for such a configuration is given by \cite{danda}:
\begin{equation}
ds^2 = g^{-\frac{1}{3}}dx_{3+1}^2 +
        g^{-\frac{1}{3}}g_{m\overline{n}}dz^mdz^{\overline{n}} +
        g^{\frac{2}{3}}\delta_{\alpha \beta}dx^{\alpha}dx^{\beta},
\end{equation}
and the 4-form field strength:
\eqr
F_{m \overline{n} \alpha \beta} & = &
        \frac{i}{4} \epsilon_{\alpha \beta \gamma} \partial_{\gamma}
                g_{m \overline{n}} \label{F_g_start} \\
F_{m89(10)} & = & -\frac{i}{2} \partial_m g \\
F_{\overline{m}89(10)} & = & \frac{i}{2} \partial_{\overline{m}} g.
\label{F_g_end}
\rqe
where $g_{m\overline{n}}$ is required to be a K\"ahler metric determined by:
\begin{equation}
\partial_{\gamma}\partial_{\gamma} g_{m\overline{n}} +
4 \partial_m\partial_{\overline{n}} g = J_{m \overline{n}}
\label{source}
\end{equation}
where $J$ is the source specifying the position of the M5-brane. The square
root of the determinant of the K\"ahler metric is denoted
$g=g_{v \overline{v}}g_{s \overline{s}}-g_{v \overline{s}}g_{s\overline{v}}$.
For later use we define a K\"ahler potential
$K(w,\overline{w},y,\overline{y},t)$ so that
\mbox{$g_{m\overline{n}} = \partial_m\partial_{\overline{n}} K$}.

\section{Decoupling limit of wrapped M5-branes}

In this section we will describe the near-horizon limit and the equations
which must be solved in this limit. This is essentially the same as the
limit taken in the special case of orthogonally intersecting M5-branes
\cite{LastPaper}. We simply take the limit where all field theory quantities
(gauge couplings and masses) are fixed while the eleven-dimensional Planck
length is taken to zero, $l_P \rightarrow 0$. We label the new coordinates,
fixed under the scaling, as in \cite{LastPaper}:
\begin{eqnarray}
w & = & \frac{v}{\alpha'} = \frac{vR}{l_{p}^{3}} \nonumber\\
t^2 & = & \frac{r}{g_s\alpha'^{\frac{3}{2}}} = \frac{r}{l_{p}^{3}} \\
y & = & \frac{s}{R}\nonumber. \label{scale}
\end{eqnarray}
where $t$ is real while $w$ and $y$ are complex.
The metric now becomes:
\begin{equation}
\frac{1}{l_{p}^2}ds^2 = g^{-\frac{1}{3}}\eta_{\mu
\nu}dx^{\mu}dx^{\nu} +
        g^{-\frac{1}{3}}g_{m\overline{n}}dz^mdz^{\overline{n}} +
        g^{\frac{2}{3}}(4t^2 dt^2 + t^4 d\Omega_{2}^2)
\label{soln_11}
\end{equation}
where now $m,n$ run over $y,w$ and $d\Omega_{2}^2$ is the metric on the round
unit 2-sphere. The source equations become:
\begin{equation}
\frac{1}{4t^5}\partial_{t}(t^3\partial_{t}) g_{m\overline{n}} +
4\partial_{m}\partial_{\overline{n}}g
 = -\pi^2\frac{\delta (t)}{t^5}\tilde{J}_{m\overline{n}}(\Sigma)
\label{metric}
\end{equation}
where $\tilde{J}_{m\overline{n}}(\Sigma)$ is the source specifying the location
of the curve $\Sigma$ in $\CC^2$. The normalisation is such that for a single
M5-brane located at $t=y=0$ we would have
$\tilde{J}_{y\overline{y}}=\delta^{(2)}(y)$. The problem is now, given some
$\Sigma$, to solve the source equations~(\ref{metric}) for the K\"ahler metric.
The holomorphic curve $\Sigma$ can be specified as the zero locus of a holomorphic
function $f(w,y)$.  So the general problem is to find the K\"ahler potential $K$
in terms of $f, \overline{f}$ and $t$.

\section{Near-horizon geometry of wrapped M5-branes}

This paper is based on a set of mathematical identities relevant 
for solving the equation (\ref{metric}).  The key observation 
is the following.  Consider a K{\" a}hler
potential consisting of two terms:
\eq
K = K^{(1)}(t,F(w,y), \overline{F}) + K^{(2)}(G(w,y), \overline{G}), \label{kahler}
\qe
where $K^{(1)}(t, F, \overline{F})$ depends on $w,y$ only through the holomorphic
function $F$, and $K^{(2)}(G, \overline{G})$ does not depend on $t$ but depends
on $w,y$ only through the holomorphic function $G$.  It is then clear 
that
the K{\" a}hler metrics derived solely from $K^{(1)}$ or
$K^{(2)}$ have vanishing determinants (since they only
depend on the variables through a single holomorphic function each).
Thus the determinant of the total metric comes from cross terms.

If we use the following ansatz for $K^{(1)}$:
\eq
K^{(1)}(t,F, \overline{F}) = \frac{c}{4t^2}\ln\frac{\sqrt{t^4+|F|^4}+t^2}
{\sqrt{t^4+|F|^4}-t^2}
\qe
we can solve the differential equation (\ref{metric}) as long
as the determinant of the metric $g$ is given by the expression:
\eq
g = \frac{c}{8(t^4+|F|^4)^{3/2}}. \label{det}
\qe

Thus to satisfy the differential equation completely
we only need to determine $K^{(2)}$ through the condition for
the determinant.  This condition results in the following equation\footnote{
In fact since the K\"ahler metric derived from $K^{(2)}$ is independent of $t$
it can be seen that its determinant must vanish to satisfy equation~(\ref{det})
and so it must only depend on a single holomorphic function as we have assumed.}:
\eq
|\partial_yF^2\partial_{w}G-\partial_{w}F^2\partial_yG|^2
\partial_G\partial_{\overline{G}}K^{(2)} = 1. \label{detcon}
\qe
To get a relation between $F$ and $G$ which respects their holomorphicity
we must have
$\partial_G\partial_{\overline{G}}K^{(2)} = |H(G)|^2$, where
$H$ is a holomorphic
function of $G$.  Hence by appropriately picking $G$ one can cast 
$K^{(2)}$ in the form:
\eq
K^{(2)} = |G(w,y)|^2,
\qe 
with $G$ determined in terms of $F$ by\footnote{On the right hand side
of this equation we can have an arbitrary phase but this can simply be absorbed
in $G$.}:
\eq
\partial_yF^2\partial_{w}G-\partial_{w}F^2\partial_yG = 1.
\label{G-cond}
\qe
We turn next
to determining $F$.

We have so far discussed satisfying (\ref{metric}) away from
any delta function singularities.  As it turns out there are 
delta functions appearing on the right hand side.  These are
localized at:
\eq
t^4 + |F|^4 =0.
\qe
Since both $t$ and $|F|$ are non-negative they must both
vanish separately at the delta function singularities.  The M5-brane
configuration we are interested in is localized at $t=0$ on a Riemann
surface.  As we are
interested in M5-branes wrapped on holomorphic curves it is
sensible to pick $F$ to be such that it vanishes on the 
holomorphic curve.  

Let $f(w,y)=0$ be the equation for the holomorphic curve
on which we wish to wrap the M5-brane.  Let the degree of
$f$ in $w$ be denoted by $N$.  Then if we normalize $f$ so that
there is no dimensionful parameter multiplying the $w^N$ term,
we see that $F$ is determined by dimensional analysis to
be
\eq
F = f^{\frac{1}{N}}.
\qe

\subsection{Sources}
To completely fix the solution we need to specify the precise form
of the sources $\tilde{J}_{m\overline{n}}(\Sigma)$. This will
allow us to determine the constant $c$
appearing in the K{\"a}hler potential.  

As we are considering a single M5-brane wrapping a Riemann surface
we have to ensure that $\tilde{J}_{m\overline{n}}(\Sigma)$ has support only on 
the
surface and is normalized such that the total M5-brane charge is
$1$.  These requirements are satisfied by
\eq
\tilde{J}_{m\overline{n}}(\Sigma) = \delta 
(f)\partial_{m}f\partial_{\overline n}\overline{f}. {\label{src}}
\qe
The condition that the M5-brane charge should be 1 is satisfied as long 
as we integrate once over the ``$f$-plane''.  

On the other hand when we plug our ansatz for the metric into
equation (\ref{metric}) there is a delta function source on the right-hand 
side of the equation which needs to be compared to the source 
(\ref{src}):
\eq
-\pi^2\frac{\delta (t)}{t^5}\tilde{J}_{m\overline{n}}(\Sigma) = 
-c\frac{\pi}{4N}\frac{\delta (t)}{t^5}\delta 
(f)\partial_{m}f\partial_{\overline n}\overline{f}.
\qe
This allows us to fix $c$ using (\ref{src}): 
\eq
c = 4\pi N.
\qe
Now the metric is completely determined with the correct normalization.

One can check that this metric agrees with the one presented in
\cite{LastPaper} when we use the specific (singular) Riemann surface:
\eq
f = w^{N}\prod_{i=1}^{n}\sinh(y-y_{i}),
\qe
which describes localized intersections of M5-branes.

\section{The example of SU($N$) Yang-Mills}



In this section we will obtain the explicit 
K{\" a}hler form for the supergravity solution 
of M5 brane wrapped
on the Seiberg-Witten Riemann surface relevant for
SU($N$) Yang-Mills \cite{AF, witn2}. 

As explained
earlier, we have to find a suitable function
$G(w,y)$ which satisfies (\ref{G-cond}) for a
Riemann surface $f(y,w) = 0$.
If we substitute the relation between
$F$ and the polynomial representing
the surface $f$ and introduce a function $h$
given by,
$G = \frac{N}{2} f^{1-\frac{2}{N}} h $ the
relation reduces to,
\eq
df \wedge dh = dw \wedge dy. \label{newG-cond}
\qe
In order to express $h$ in a concise manner
let us change the independent set of variables
from $(y,w)$ to $(f,w)$ . After making this 
substitution in the above equation 
(\ref{newG-cond}), the equation for $h(f,w)$
becomes
\eq
\partial_w h(f,w) = -\partial_f y  \label{h-eq1},
\qe
while the $\partial_f h(f,w)$ remains undetermined.
This is due to the ambiguity in splitting the 
K{\"a}hler form and does not change the metric. 
For a suitable choice of $f$ this equation can be
integrated to obtain a solution for $h$ as
\eq
h(f,w) = - \int {dw} (\partial y /\partial f) (f,w)
\label{h-eq}
\qe

Let us consider an ${\cal N}=2$ gauge theory with 
gauge group SU($N$) , 
corresponding to a pair of parallel 
NS5 brane with $N$ D4 branes stretched between them \cite{witn2}.
The M-theory lift of this configuration, as mentioned
in section 2, will be
an M5 brane wrapped on a Riemann surface $\Sigma$
embedded holomorphically in $\CC^2$ with complex
coordinates $w$ and $y$ (\ref{scale}).
The Riemann surface is given by the holomorphic equation:
\eq
f = e^{y} + 2 B(w) + e^{-y} = 0\label{surface}
\qe
where $B(w)$ is a general polynomial in $w$ of degree $N$.
As is well-known the moduli space of this gauge 
theory is the same as that of the associated Riemann surface.

There is a direct connection
between the polynomial $B(w)$ and the gauge theory as described 
in \cite{AF}. 
The parameters of the gauge theory moduli
space  $s_\alpha$ can be expressed in terms of the Higgs VEV
$a_I$ in the Cartan subalgebra through the relation
\begin{equation}
s_\alpha = (-)^\alpha 
\sum_{I_1<I_2<...<I_\alpha} a_{I_1}...a_{I_\alpha}.
\end{equation}
These parameters occur as the coefficients in the polynomial
$B(w)$ as,
\begin{equation}
B(w) = \sum_{\alpha=0}^n s_\alpha w^{n-\alpha}. 
\label{curve}
\end{equation}

The function $h$ for the present configuration
can be obtained by integrating (\ref{h-eq}) for
the surface given by (\ref{surface}).
Using $\cosh{y} = \sqrt{(\frac{f}{2} -B(w))^2-1}$ we can write down
the solution 
\eq
h(f,w) = -(1/2) \int_0^w \frac{dw}{\sqrt{(\frac{f}{2}-B(w))^2-1}}
\qe
which can be expressed in a parametric form
\eq
h(f,w) = -(w/2) \int_0^1 \frac{dt}{\sqrt{(\frac{f}{2}-B(tw))^2-1}}. \label{para}
\qe

While we have only worked out the SU($N$) Yang-Mills case explicitly, 
in principle the function $G$ can be found for any Seiberg-Witten curve.
All one has to do is solve equation (\ref{G-cond}).  


\section{Conclusions and discussion}
In this paper we have presented new solutions of 11-dimensional 
supergravity.  These solutions represent M5-branes wrapped on Riemann
surfaces embedded in 4-dimensional space.  Our solutions preserve 
8 real supersymmetries and are dual to ${\cal N}=2$ Seiberg-Witten theories
through Maldacena's conjecture.  

Our approach allows one, in principle, to find the supergravity dual
for {\em any} Seiberg-Witten theory.  In this sense, our approach
may be used to classify geometries dual to ${\cal N}=2$ gauge 
theories\footnote{Another approach to the problem of M5-branes wrapping
Riemann surfaces was presented in \cite{maldanunez}.}.

Although our approach yields the geometry and relevant supergravity
quantities we have not tried here to relate them to field theory 
quantities such as coupling constants, nor have we made any serious 
attempt at studying the physics of these theories using the supergravity solution. 
We leave these issues for future work.

One future direction would be to study specific interesting models such as
SU($N$) Yang-Mills theory, whose supergravity solution is presented in 
our paper.  It would be interesting to compare the large $N$ behaviour of 
the supergravity solution
to the field theory analysis of Douglas and Shenker\cite{ds}.  
Another question concerns renormalization group flow in these 
theories, which may be
richer and, hopefully, more manageable than the corresponding problem in type IIB
theory with D3-branes.  Here one would study the flow of Riemann surfaces
into the infrared and determine the corresponding behaviour in the 
dual field theory.  

We hope that our approach will broaden the class of theories amenable 
to study using Maldacena's conjecture.

\section{Acknowledgements}
The research of AF and SM is funded by the Swedish Research
Council (NFR).

\end{document}